\documentclass[%
 reprint,
superscriptaddress,
 amsmath,amssymb,
 aps,
prb,
longbibliography
]{revtex4-2}

\usepackage{graphicx}%
\usepackage{dcolumn}%
\usepackage{multirow}
\usepackage{amssymb}
\usepackage{comment}
\usepackage{MnSymbol,wasysym}
\usepackage{setspace}

\usepackage[dvipsnames,table,xcdraw]{xcolor}

\usepackage{hyperref}
\hypersetup{
    colorlinks=true,
    linkcolor=Blue,
    filecolor=Blue,
    urlcolor=Blue,
    citecolor=blue,
}

\renewcommand\vec[1]{\ensuremath\boldsymbol{#1}} 
\newcommand{\D}{\mathrm{d}} 

\begin{document}
\title{Entanglement signatures of topological phase transitions in a dirty Weyl semimetal}

\author{Vin\'icius Mohr}
\affiliation{Institute for Theoretical Physics, ETH Zurich, 8093 Zurich, Switzerland}

\author{Shunji Tsuchiya}
\affiliation{Department of Physics, Chuo University, 1-13-27 Kasuga, Bunkyo-ku, Tokyo 112-8551, Japan}

\author{Andr\'as L. Szab\'o}
\affiliation{Institute for Theoretical Physics, ETH Zurich, 8093 Zurich, Switzerland}
\affiliation{Max Planck Institute for Solid State Research, D-70569 Stuttgart, Germany}
\date{\today}

\begin{abstract}
Three-dimensional Weyl semimetals (WSMs) constitute paradigmatic gapless topological phases whose nodal structure is stable against weak perturbations, including disorder.
Two distinct ways to destroy the topology of the Weyl nodes are through pairwise annihilation in momentum space or by sufficiently strong disorder that eliminates the quasiparticle pole, driving the system into a non-Fermi-liquid diffusive-metallic state.
In this work, we study the evolution of entanglement spectrum and entanglement entropy of a dirty WSM as it undergoes these two topological transitions.
In the clean limit, the WSM topology is reflected by a locus of $\xi=1/2$ eigenvalues of the reduced correlation matrix mirroring the structure of Fermi arcs. 
We show that disorder broadens this feature into a finite-width distribution, which disappears either through a gradual loss of spectral weight upon node annihilation or by melting into the background at the onset of metallicity.
When tuning across the disorder-driven transition, the scaling of the R\'enyi entropies observed for weak disorder gradually breaks down as the system approaches the critical disorder strength.

\end{abstract}

\maketitle

\section{Introduction}

Topology characterizes global properties of a system that remain invariant under smooth deformations, endowing topological phases with robustness against weak perturbations, including disorder.
While topological insulators are naturally protected by a bulk energy gap, in gapless topological phases, such as in Weyl semimetals (WSMs), this stability is rooted in the momentum-space separation and topological charge of nodal points~\cite{Yan2017,Hasan2017,Armitage2018}.
Probing nonlocal characteristics of quantum states, entanglement-based measures emerged as powerful probes of topological phases.
In interacting systems, the scaling of entanglement entropy (EE) with subsystem size has provided important insight into phases with long-range topological order~\cite{Kitaev2006}.
On the other hand, in free-fermion systems, the entanglement spectrum (ES)~\cite{LiHaldane2008} has been shown to encode boundary physics, including chiral edge modes, surface states in topological insulators and superconductors~\cite{Fidkowski2010,Turner2010,Hughes2011,Alexandradinata2011}, as well as the Fermi arcs in WSMs~\cite{Wang2014,Zhou2023}.

In the presence of disorder, the loss of translational invariance complicates the study of topological phases by precluding the usual momentum-space formulation of invariants.
EE between subsystems defined through real-space partitioning is therefore particularly well-suited to studying disordered systems.
In this context, both EE and ES have been shown to capture topological transitions in the presence of disorder in a one-dimensional Su-Schrieffer-Hegel chain~\cite{Kumar2026}, as well as two-dimensional Chern~\cite{Prodan2010} and quantum-spin Hall~\cite{Gilbert2012} insulators.

\begin{figure}
    \begin{center}
	\includegraphics[width=0.98\linewidth]{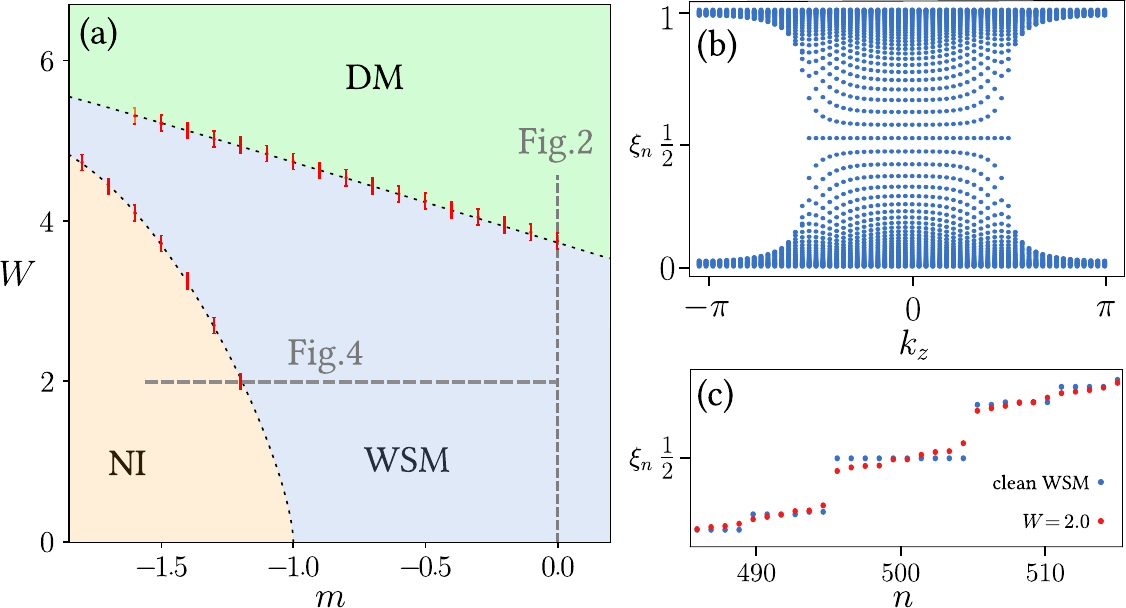}
    \end{center}
    \caption{(a) Phase diagram of the Hamiltonian in Eq.~(\ref{eq:Hamiltonian}) in the presence of disorder with strength $W$, obtained from reconstructing the average density of states.
    Both $m$ and $W$ are measured in units of $t$.
    For details of the computation see Appendix~\ref{sec:A}.
    (b) Momentum-resolved ES in the clean system for $L=60$ linear dimension.
    (c) Sorted eigenvalues $\xi_n$ with index $n$ from the middle of the ES for $W=0$ (blue) and $W=2.0$ for one specific disorder realization (red).}
    \label{fig:PhaseDiagram}
\end{figure}

\begin{figure*}[t!]
    \centering
    \includegraphics[width=0.95\linewidth]{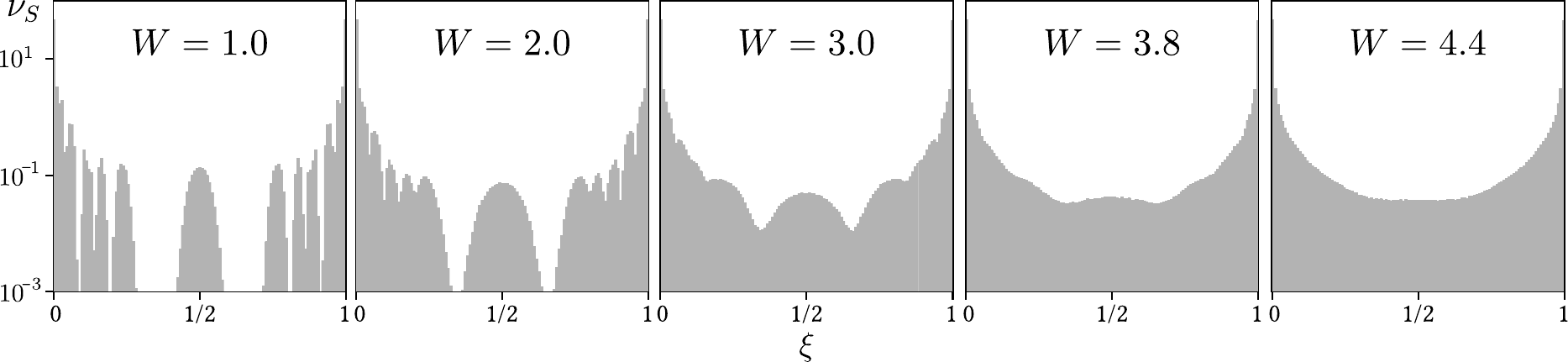}
    \caption{Evolution of entanglement density of states (EDOS) for increasing disorder at $m=0$. The peak at $\xi=1/2$ gradually broadens in the WSM phase until it melts into the background near the critical disorder strength $W_{\rm c}\approx 3.75t$.}
    \label{fig:EDOS-W}
\end{figure*}

Here, we address a minimal gapless three-dimensional topological WSM, characterized by a single pair of Weyl nodes in the bulk band structure, and subject to random potential disorder.
The nodes act as sources and sinks of Berry curvature, and are connected by a locus of topological surface states forming the Fermi arcs.
Remarkably, the WSM remains stable against weak disorder, while sufficiently strong disorder destroys the quasiparticle pole at a critical disorder strength and drives the system into a non-Fermi-liquid diffusive-metallic phase~\cite{Sbierski2014,Chen2015,Syzranov2015,Louvet2016,Roy2018,Shapourian2016,Ohtsuki2020}.
The ensuing compressible metal is topologically trivial. The destruction of Weyl nodes is accompanied by a gradual dissolution of Fermi arc states, which lose their surface localization~\cite{Slager2017}.
This semimetal-to-metal transition is controlled by a disorder-induced, or ``dirty'' quantum-critical point, which has been extensively studied~\cite{Fradkin1986,Ominato2014,Altland2015,Sbierski2015,Roy2016,Ohtsuki2018,Syzranov2016, Kobayashi2014,Pixley2016b,Bera2016,Fu2017}. An alternative perspective instead attributes the transition to avoided criticality arising from rare-region effects~\cite{Nandkishore2014,Pixley2016,Pixley2021}.
We note that upon further increasing disorder strength, the diffusive metal (DM) eventually undergoes Anderson localization~\cite{Pixley2015,Pixley2016c}.
However, in this work we are not concerned with the localization transition.

Even though both the WSM and DM represent gapless phases of matter, the former features a codimension three (zero-dimensional) Fermi point, whereas the latter is devoid of a sharp Fermi surface altogether. As a result, the EE in both phases obeys area law, free of the logarithmic corrections that a conventional Fermi surface would produce~\cite{Li2006,Schollwoeck2006,Swingle2010,Potter2014}.
Nevertheless, the ES retains a distinct signature of the topological character of the WSM:
Analogously to surface states in the presence of physical boundaries, the topological obstruction to the atomic limit produces a set of $\xi=1/2$ eigenvalues of the reduced correlation matrix~\cite{Fidkowski2010,Turner2010,Hughes2011,Wang2014,Zhou2023}, which connect the Weyl nodes as they are projected onto the virtual boundary.

In this work, we explore the evolution of EE and ES as the topological band structure of the WSM is destroyed in two fundamentally distinct ways: by increasing disorder strength past its critical value, such that the quasiparticle pole collapses, driving the system into the DM phase; as well as by annihilating the Weyl nodes at finite but subcritical disorder via the tuning of band parameters, giving rise to a normal insulator (NI).
We find that for finite disorder, the set of eigenvalues pinned to $\xi=1/2$ in the clean limit is replaced by a distribution of finite width around 1/2, and introduce the entanglement density of states (EDOS) to track their evolution across the two distinct phase transitions.
Furthermore, we compute the disorder-averaged R\'enyi entropies of different orders, and investigate their scaling with disorder strength.
Our results demonstrate that entanglement measures remain sensitive to distinct topological transitions occurring in a gapless three-dimensional WSM.

\section{Model}

We study a minimal, two-orbital lattice model, constructed by stacking two-dimensional topological insulators in momentum space.
The corresponding tight-binding Hamiltonian is of the form~\cite{Kobayashi2014,Ohtsuki2016,Shapourian2016,Chen2015}
\begin{equation}
H_0 = \sum_{\boldsymbol{k}} \psi_{\boldsymbol{k}}^{\dagger} \big[\boldsymbol{N}(\boldsymbol{k}) \cdot \boldsymbol{\sigma}\big] \psi_{\boldsymbol{k}},\label{eq:Hamiltonian}
\end{equation}
where $\psi_{\vec{k}}=(c_{1,\vec{k}}, c_{2,\vec{k}})^\top$, and $c_{i,\vec{k}}$ annihilates a fermion with momentum $\vec{k}$ and orbital $i=1,2$.
Here, $\boldsymbol{\sigma}=(\sigma_1,\sigma_2,\sigma_3)$ are Pauli matrices acting in orbital space, furthermore
\begin{align}
    &N_{1}(\vec{k})=t \sin(k_{x} a), \hspace{0.3cm} N_{2}(\vec{k})=t \sin(k_{y} a), \nonumber\\
    &N_3(\vec{k})=2-m-t_z \cos(k_z a)-t_0\sum_{i=x,y}^2\cos(k_i a),
\end{align}
yielding the band energies $E_{\vec{k}}=\pm \sqrt{N_1^2+N_2^2+N_3^2}$, and we implicitly set the chemical potential to zero.
For $t_z=0$, the above Hamiltonian describes a set of two-dimensional Chern insulators in the $(k_x,k_y)$ plane, with the mass $m$ tuning the system through phases with distinct Chern numbers via topological gap closing and reopening.
Finite $t_z$ introduces interlayer hopping, and effectively modulates $m$ along the $k_z$ axis.
We set $t_z/t=t_0/t=1$, the lattice constant $a=1$, for which the system hosts one (two) pair(s) of Weyl nodes for $-1<m/t<1$ and $3<m/t<5$ ($1<m/t<3$), and a NI phase otherwise.

For concreteness, in this work we focus on the region $m\leq 0$.
Then, for $m/t>-1$ the gap closes at two singular points $\vec{k}=(0,0,\pm \arccos m/t)$, giving rise to the Weyl nodes and representing topological gap closing in the constituent two-dimensional insulators.
Accordingly, the system has a nonvanishing Chern number between the Weyl nodes, forming the Fermi arcs in the $(k_x,k_z)$ or $(k_y,k_z)$ planes.
In real space, the arcs constitute chiral edge states localized to the $(x,z)$ and $(y,z)$ surfaces.
Tuning $m/t\to-1^+$, the Weyl points approach each other and annihilate, leading to the topologically trivial NI.
We remark that for $|t_z|/t<1$, the model in Eq.~(\ref{eq:Hamiltonian}) also hosts a three-dimensional topological insulator (TI) phase, devoid of Weyl nodes.
Similarly to the WSM, this phase is also driven into the DM by strong disorder~\cite{Chen2015,Roy2018,Shapourian2016,Ohtsuki2018,Kobayashi2014}.
Here, however, we do not delve into this phase.

Disorder is introduced as an orbital-independent on-site potential $V(\vec{r})$, drawn independently for each lattice site from a uniform random distribution $[-W/2,W/2]$, where the disorder strength $W$ is measured in units of $t$.
In the WSM, weak disorder introduces corrections to both the real and imaginary parts of the self-energy, yielding renormalized quasiparticles with an energy-dependent broadening~\cite{Klier2019,Ohtsuki2020,Ohtsuki2016,Ominato2014}.
At a critical disorder strength $W_{\rm c}$, the quasiparticle pole eventually disappears, signaling 
the loss of a well-defined nodal band structure and the onset of the DM.

To anchor our investigation, we first reconstruct the phase diagram in the $(m,W)$ plane by computing the average density of states (DOS) $\nu_{\rm A}$ in a cubic system with $L=100$ linear dimension, using the kernel polynomial method~\cite{Weisse2006}, see Fig.~\ref{fig:PhaseDiagram}.
While the WSM is characterized by a qudratically vanishing DOS, $\nu_A(E)\sim |E|^2$, the NI features a finite gap around $E=0$.
In contrast, the compressible DM shows finite DOS at zero energy $\nu_A(E\approx 0)\sim{\rm constant}$.
For details and extended data see Appendix~\ref{sec:A}.

\begin{figure}
    \centering
    \includegraphics[width=0.99\linewidth]{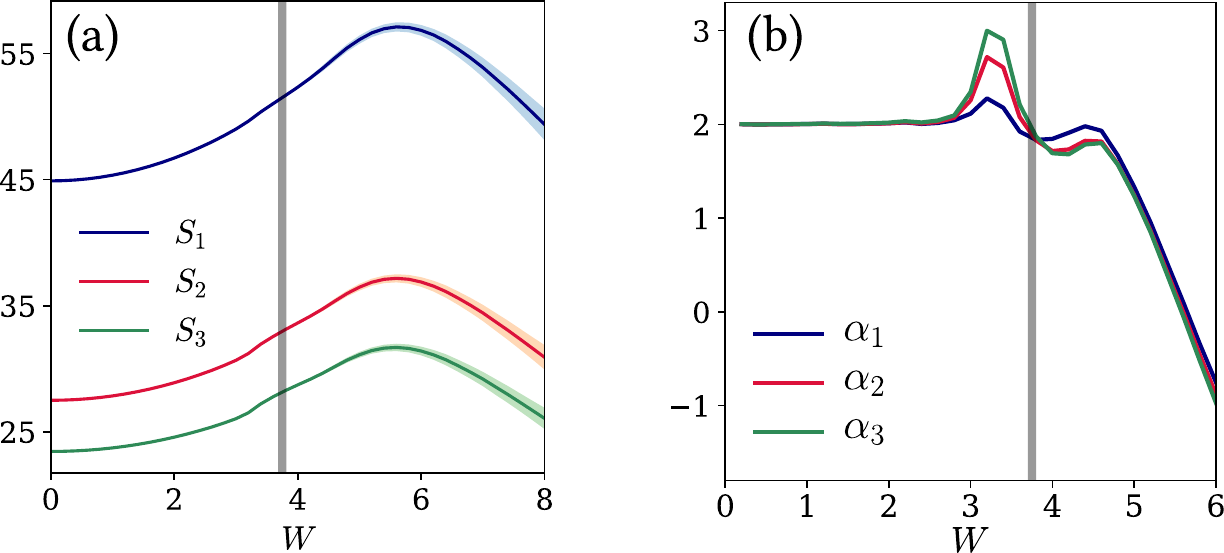}
    \caption{Entanglement measures as a function of $W$ (measured in units of $t$), with gray stripe marking $W_{\rm c}$. (a) Disorder-averaged entropies and their variance shown in the shaded area. (b) Local power-law exponent $\alpha_i$ of $\Delta S_i$.}
    \label{fig:EE}
\end{figure}

\section{Entanglement entropy and entanglement spectrum}

\subsection{Clean Weyl semimetal}

We proceed by computing the ES for the Hamiltonian in Eq.~(\ref{eq:Hamiltonian}).
We bipartition the system in real space into two equal subsystems $A$ and $B$, with the virtual cut perpendicular to the $x$ axis.
It is convenient to introduce the entanglement Hamiltonian $H_{\rm E}$, such that the reduced density matrix of subsystem $A$ is $\rho_A\propto e^{-H_{\rm E}}$.
For free fermions, the single-particle entanglement energies $\varepsilon_{\rm E}$ (the eigenvalues of $H_{\rm E}$) can be obtained from the reduced correlation matrix $C_{ij}=\langle c_i^{\dagger} c_j\rangle$, where $i,j\in A$~\cite{Peschel2003,Peschel2004}.
Then, $\varepsilon_n = \log[(1-\xi_n)/\xi_n]$, where $\varepsilon_n$ and $0\le\xi_n\le 1$ are the eigenvalues of $H_{\rm E}$ and $C_{ij}$ respectively, the latter of which span the ES.
The $i$th R\'enyi entropy for $i>0$ can be computed as
\begin{align}
    S_i = \frac{1}{1-i}\sum_n \log \big[ \xi_n^i +(1-\xi_n)^i \big],\label{eq:Si}
\end{align}
with the von Neumann entropy following from the limit $i\to1$ as
\begin{equation}
S_1=-\sum_n\Big[\xi_n\log \xi_n + (1-\xi_n)\log(1-\xi_n)\Big].\label{eq:S1}
\end{equation}
Note that for any $S_i$, the values $\xi=0,1$ contribute zero entropy and $\xi=1/2$ contributes the maximal value $\log 2$, whereas all other $\xi$ are weighted differently for different $i$.

We impose periodic boundary conditions in all three directions and compute the eigenvalues of $C_{ij}$ using exact diagonalization for $L=10$ at zero temperature, where the ground state can be represented by a Slater determinant.
In the clean limit, translational invariance is preserved in the $(y,z)$ plane, which contains the virtual boundary between subsystems.
Consequently, $\xi_n$ can be labeled by $(k_y,k_z)$, which allows us to resolve the imprint of Fermi arc states in the ES.
These manifest as a set of $\xi=1/2$ eigenvalues at $k_y=0$, connecting the Weyl nodes between $k_z=\pm \arccos m/t$~\cite{Wang2014,Zhou2023}, see Fig.~\ref{fig:PhaseDiagram}(b).
Analogously to the Fermi arcs in the physical spectrum, these are $\varepsilon=0$ states of $H_{\rm E}$, and they arise from the constituent two-dimensional Chern insulators having a topological gap between the Weyl nodes, while being trivial outside of them.
The topological obstruction to the atomic limit results in states that maximally entangle the subsystems, yielding $\log 2$ entropy.
In contrast, the rest of the ES is smoothly connected to $\xi=1$ or $\xi=0$, which respectively represent states fully localized within subsystem $A$ or $B$.
As the Weyl nodes approach each other for decreasing $m<0$, the number of $1/2$ states decreases accordingly, disappearing when the Weyl nodes annihilate at $m=-1$.

\begin{figure*}[t!]
    \centering
    \includegraphics[width=0.95\linewidth]{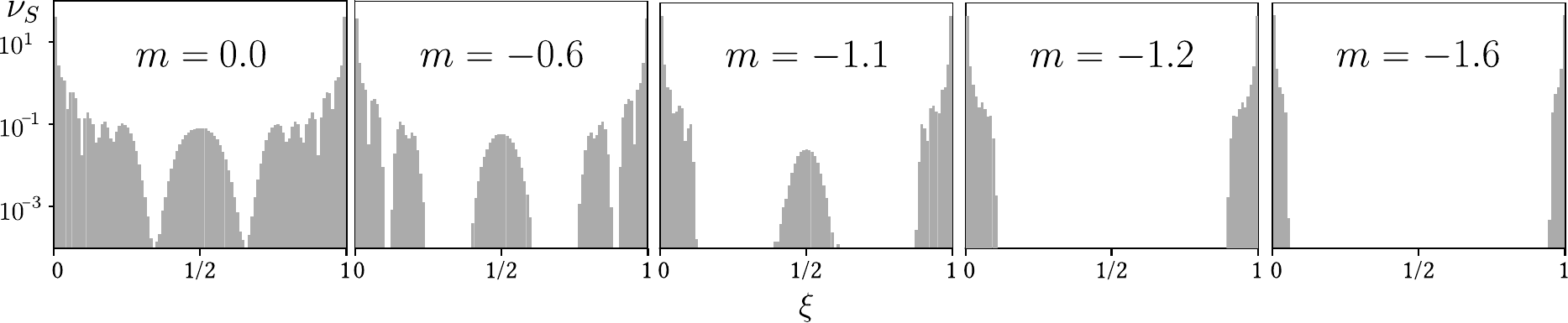}
    \caption{Evolution of EDOS at fixed $W=2.0$ upon tuning $m$ across the WSM-NI phase boundary. The peak at $\xi=1/2$ gradually loses spectral weight, and disappears at the transition point $m\approx-1.2t$.}\label{fig:WSM-NI}
\end{figure*}

\subsection{Disordered Weyl semimetal}

We proceed by investigating the evolution of the ES upon increasing $W$ starting from the clean system past the disorder-driven critical point.
For this, we set $m=0$, for which the analysis of DOS indicates $W_{\rm c}/t \approx3.75 \pm 0.1$.
We find that despite the stability of Weyl nodes for weak disorder, the $1/2$ entanglement eigenvalues are no longer pinned for $W\neq 0$, which we show in Fig~\ref{fig:PhaseDiagram}(c) for $W=2.0$.
Therefore, to capture the evolution of the ES in the absence of translational invariance, we compute the entanglement spectrum for $10\,000$ disorder realizations at each fixed $W$, and reconstruct the distribution of entanglement eigenvalues $\xi$ in the disorder ensemble, which we refer to as the EDOS or $\nu_S(\xi)$.
The evolution of $\nu_S$ as a function of $W$ is shown in Fig.~\ref{fig:EDOS-W}.
For small but finite disorder, the otherwise $\delta$-like peak at $\xi=1/2$ is replaced by a finite-width peak, which is still separated from the rest of the spectrum.
At moderate $W$ the broadened peak connects with the continuum of eigenvalues, but it remains clearly identifiable, until at $W\approx W_{\rm c}$ the peak fully loses coherence and disappears into the background.
The evolution of $\nu_S$ therefore reflects the gradual dissolution of Fermi arcs, which were found to lose surface localization for weak disorder, and melt into the bulk by the time $W_{\rm c}$ is reached~\cite{Slager2017}.

While the full ES generally contains more information than the entanglement or R\'enyi entropies, the latter provide convenient scalar quantities that admit straightforward disorder averaging.
To investigate their evolution, we compute $S_i$ for $i=1,2,3$ using Eqs.~(\ref{eq:Si}) and (\ref{eq:S1}) for each disorder realization, and average over the ensemble for fixed $W$.
We find that $S_{1,2,3}(W)$ increase monotonically throughout the WSM phase, and peak deep in the DM near $W= 5.5t$, after which they decrease, potentially as a precursor to Anderson localization at strong disorder, see Fig.~\ref{fig:EE}(a).

Notice that the entropies evolve smoothly across $W_{\rm c}$, without a clear anomaly.
Therefore, to gain further insight, we compute the logarithmic derivative via $\alpha_i(W)=(\D \ln \Delta S_i)/(\D \ln W)$, where in $\Delta S_i(W)= S_i(W)-S_i(0)$ we subtract the clean entropy.
Then,  $\alpha_i$ captures the dominant local power-law scaling of $\Delta S_i$.
We find that the WSM is characterized by quadratic scaling of $\Delta S_{1,2,3}$ with $W$ up to $W\approx 3.0 t$, which can be understood as follows.
The reduced correlation matrix is built from single-particle states, which admit a perturbative expansion for weak disorder.
As the distribution of the onsite potential $V$ has vanishing mean, the leading disorder-averaged correction is proportional to $W^2$.
For small values of $i$ and weak disorder, $\Delta S_i(W)$ is consistent with the leading-order correction.
In the neighborhood of $W_{\rm c}$, $\alpha_i$ deviates from two, indicating the increasing importance of higher-order corrections in disorder.
Power-law scaling of $\Delta S_i(W)$ eventually fully breaks down in the DM.
Nevertheless, caution is advertised upon interpreting such scaling, as the kernel in Eq.~(\ref{eq:S1}) is nonanalytic at $\xi=0,1$. 
We find that for moderate $i\lesssim 30$, $\alpha_i(W)$ behaves qualitatively similarly albeit with $\alpha_i>2$ for small $W$, whereas even higher order $\Delta S_i$ do not show power-law scaling for any $W$.

Finally, we address the scenario where the topological WSM is destroyed via the annihilation of Weyl nodes at finite disorder strength.
To this end, we set $W=2.0t$ and track the evolution of $\nu_S$ as we tune $m$ across the WSM-NI phase boundary at $m/t\approx-1.2$, see trajectory in Fig.\ref{fig:PhaseDiagram}(a).
As $m$ is decreased and the Weyl nodes are brought closer together, the peak at $1/2$, now of finite width, loses spectral weight, mirroring the shortening of Fermi arcs, see Fig.~\ref{fig:WSM-NI}.
The peak finally vanishes as the system becomes a trivial insulator.
This is in contrast with the disorder-induced transition, where the peak disappears via broadening and dissolving into the background.

In conclusion, we showed that while the topological WSM is stable for weak disorder, the eigenvalues of the reduced correlation matrix, which mirror the structure of physical surface states, do not remain pinned and deviate from $\xi=1/2$.
Nevertheless, the ES overall continues to reflect the presence of topological band structure, which we captured by tracking the evolution of EDOS, showing a broadened peak centered at $\xi=1/2$.
Removing the Weyl nodes via the destruction of quasiparticles at the disorder-induced WSM-DM transition broadens the peak until it fully loses coherence and disappears into the background.
In this context, we also showed that the lowest-order R\'enyi entropies exhibit $\sim W^2$ scaling for weak disorder, which breaks down upon approaching the dirty quantum critical point.
Finally, the annihilation of the nodes via the tuning of the mass parameter $m$ across the WSM-NI transition eliminates the peak by suppressing its spectral weight.
Altogether, our results show that entanglement measures distinguish between disorder- and band-structure-driven topological transitions in gapless systems.

\emph{Acknowledgments.}
We thank S. Haas, J. Budich, and S. Qazim for enlightening conversations. A. S. was supported by the Simons
Foundation (Grant number SFI-MPS-NFS-00006741-11). S.T. is supported by JSPS KAKENHI Grant Number JP25K07158.

\appendix

\section{Extended data}\label{sec:A}

\begin{figure*}[t]
    \centering
    \includegraphics[width=0.8\linewidth]{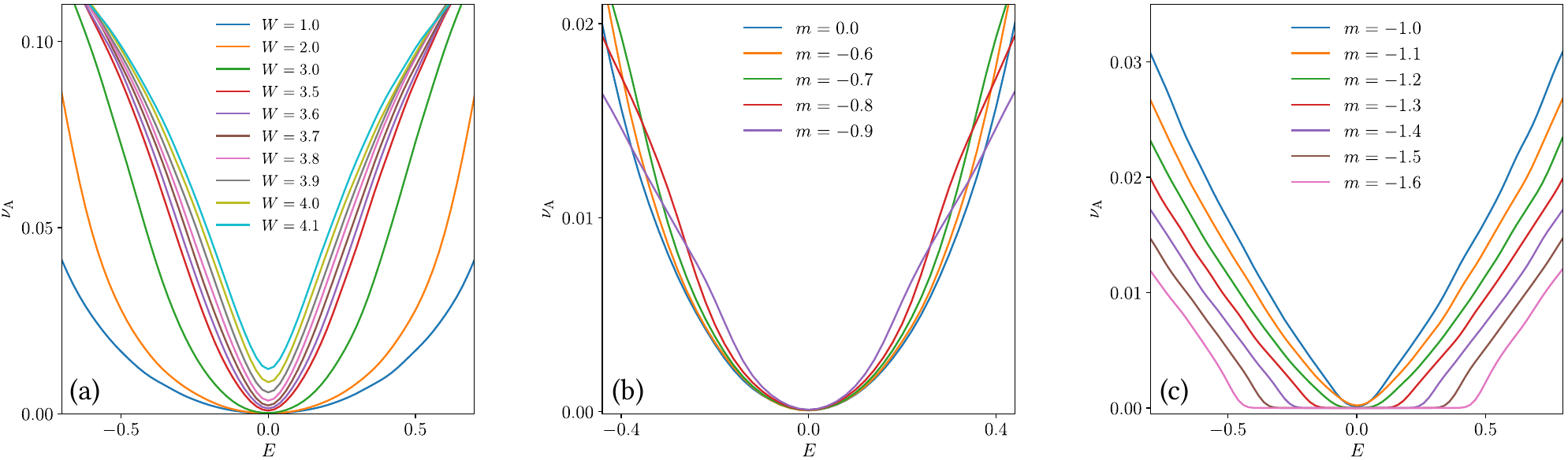}
    \caption{Density of states for fixed (a) $m=0.0$ and (b)--(c) $W=2.0$, corresponding to the gray lines in the $(m,W)$ plane in Fig.~\ref{fig:PhaseDiagram}(a) and the respective data sets in Figs.~\ref{fig:EDOS-W} and \ref{fig:WSM-NI} of the main text }
    \label{fig:KPM}
\end{figure*}

\begin{figure}[t!]
    \centering
    \includegraphics[width=0.95\linewidth]{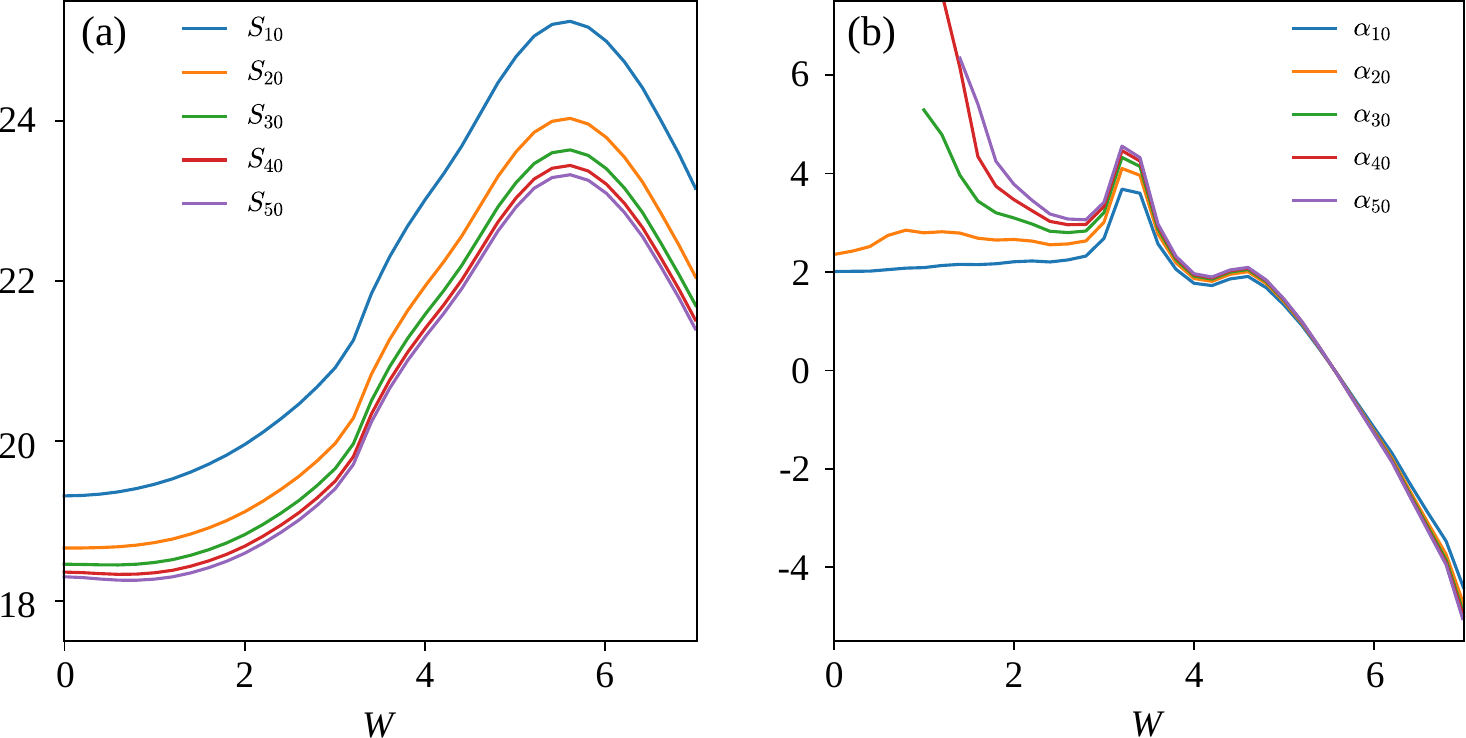}
    \caption{(a) R\'enyi entropies of order 10, 20, 30, 40, 50 and (b) their local power-law exponent, as a function of disorder strength (in units of $t$).}
    \label{fig:Extended_SData}
\end{figure}

\emph{Phase diagram.}
We first present details of the computation of the phase diagram in Fig.~\ref{fig:PhaseDiagram}(a), alongside further data sets.
To obtain the phase boundaries between the WSM, DM, and NI phases, we reconstruct the average DOS $\nu_{\rm A}(E)$ on cubic lattice with linear system size $L=100$ on a grid in the $(m,W)$ plane with step size $\Delta m=\Delta W =0.1t$.
To this end, we expand $\nu_{\rm A}(E)$ on Chebyshev polynomials and compute the first 1000 weights using the kernel polynomial method~\cite{Weisse2006}.

While the WSM is characterized by $\nu_{\rm A}(E)\sim |E|^2$, the NI shows a finite spectral gap, whereas in the DM, $\nu_{\rm A}(E=0)$ becomes finite, see Fig.~\ref{fig:KPM}, where we show data sets corresponding to Figs.~\ref{fig:EDOS-W} and~\ref{fig:WSM-NI}.
The error bars in Fig.~\ref{fig:PhaseDiagram} are taken conservatively as one full step $\Delta W$.

\emph{R\'enyi entropies}
Next, we present extended data for higher-order R\'enyi entropies.
We use the same ensemble of entanglement eigenvalues $\xi_n$ as for Fig.~\ref{fig:EDOS-W}, with 10\,000 disorder realizations for each $W$.
Figure~\ref{fig:Extended_SData} shows R\'enyi entropies of order 10, 20, 30, 40, and 50 and their respective local power-law exponents.

\bibliography{bibliography}
\end{document}